\documentclass[a4paper,11pt]{article}
\pdfoutput=1 

\usepackage{jinstpub} 

\usepackage{comment}
\usepackage{graphicx}
\usepackage{adjustbox}
\usepackage{notoccite}
\usepackage{hyperref}

\usepackage{caption}
\usepackage{subcaption}

\usepackage{mfirstuc}

\newcommand{\addComment}[2]{
  \expandafter\newcommand\csname #1\endcsname[1]{{\bf \color{#2} \capitalisewords{#1}:\,##1}}
  \expandafter\newcommand\csname #1cor\endcsname[2]{{\color{#2} \capitalisewords{#1}:\,\st{##1}{\bf ##2}}}
  \expandafter\newcommand\csname #1color\endcsname{#2}
}

\addComment{cris}{blue}

\title{Design of Detectors at the Electron Ion Collider with Artificial Intelligence}

\author[a,*]{C. Fanelli}

\affiliation[a]{Laboratory for Nuclear Science, Massachusetts Institute of Technology, Cambridge, MA 02139, U.S.A.}

\emailAdd{cfanelli@mit.edu}

\abstract{Artificial Intelligence (AI) for design is a relatively new but active area of research across many disciplines. Surprisingly when it comes to designing detectors with AI this is an area at its infancy. 
The Electron Ion Collider is the ultimate machine to study the strong force. The EIC is a large-scale experiment with an integrated detector that extends for about $\pm$35 meters to include the central, far-forward, and far-backward regions. 
The design of the central detector is made by multiple sub-detectors, each in principle characterized by a multidimensional design space and multiple design criteria also called objectives. 
Simulations with Geant4 are typically compute intensive, and the optimization of the detector design may include non-differentiable terms as well as noisy objectives. 
In this context, AI can offer state of the art solutions to solve complex combinatorial problems in an efficient way. 
In particular, one of the proto-collaborations, ECCE, has explored during the detector proposal the possibility of using multi-objective optimization to design the tracking system of the EIC detector. 
This document provides an overview of these techniques and recent progress made during the EIC detector proposal.
Future high energy nuclear physics experiments can leverage AI-based strategies to design more efficient detectors by optimizing their performance driven by physics criteria and minimizing costs for their realization.
}

\keywords{{\color{blue} AI/ML, detector design, bayesian, evolutionary, genetic, multi-objective optimization}}

\arxivnumber{} 

\proceeding{1$^{\text{st}}$ Workshop on Artificial Intelligence for the Electron Ion Collider:\\ Experimental applications\\
  September 7-10\\
  Center for Frontiers in Nuclear Science}

\begin{document}
\maketitle
\flushbottom

\section{Introduction}
\label{sec:intro}

Design assisted by Artificial Intelligence is a relatively new but active area of research across multiple domains.\footnote{In this proceeding we will follow a common taxonomy \cite{cfanelli_ai4eic_intro} according to which AI is a broad set of algorithms/methods that encompass Machine Learning (ML), which in turn contains the sub-domain of Deep Learning (DL) consisting of multi-layered deep neural networks (DNN).} 
In industrial material, for example, a breadth of different AI-based methods has been applied for material design spanning from evolutionary algorithms for the design of modular metamaterials to Graph Neural Networks for hardness prediction and architected material design (see \cite{guo2021artificial} for a more detailed discussion on the design with AI/ML of mechanical materials). 

Another active field of research is that of molecular and drug design: for pharmacologically relevant small molecules, the number of theoretically possible structures is estimated to be $\mathcal{O}$(10$^{60}$) of which perhaps only a billion are realizable \cite{virshup2013stochastic}. 
Simulation of course offers a way of probing this space without experimentation, nevertheless the size of chemical space is still overwhelming, and a `smart navigation' is required: applications range from optimization of molecules via Deep Reinforcement Learning \cite{zhou2019optimization} to inverse molecular design using generative models for discovery of new materials \cite{sanchez2018inverse}. 
Inverse design in particular is an appealing approach that can accelerate the design process in that starts from the desired properties (also called in what follows objective space) and ends in the design space, differently from the direct approach that leads to the objective space from the design space. 
As discussed in \cite{sanchez2018inverse}, the combination of inverse design methods with active learning approaches like, \textit{e.g.}, Bayesian optimization \cite{snoek2012practical}
can allow a model to adapt while exploring the design space, promoting the discovery of regions with desirable properties. 
Automated recommendation tools that leverage machine learning and probabilistic modeling techniques have been developed to guide synthetic biology \cite{radivojevic2020machine}; many other fields benefit from an AI-supported design, see, \textit{e.g.}, \cite{so2020deep}.

When it comes to designing detectors with AI this is an area at its `infancy'. 
Fundamental nuclear and particle physics research often requires constructing large-scale experiments that combine multiple sub-detectors to investigate the building blocks of nature.
 Each sub-detector is typically characterized by multiple parameters that control \textit{e.g.}, the geometry, the mechanics and the optics of each component. 
Traditionally each sub-detector is first studied individually taking into account the constraints from the baseline full detector; the global detector is eventually studied and characterized after all the sub-detector systems are ready. 
The global detector design can involve a high-dimensional space which consequently implies a well-known phenomenon called ``curse of dimensionality'' \cite{bellman1957dynamic}, corresponding to a combinatorial explosion of possible values to search. 
In addition to a multidimensional \textit{design space} we may also need to consider multiple objective functions in our \textit{objective space} to characterize the design of a detecting system. 
Things are further complicated by the fact that we need accurate but computationally intensive simulations using toolkits like Geant4 \cite{agostinelli2003geant4} to simulate the passage of particles through matter using Monte Carlo methods. 


In this context, AI offers state of the art solutions spanning from evolutionary algorithms to reinforcement learning in order to solve complex optimization problems in an efficient way by reducing the computing budget needed to converge to optimal solutions (see, \textit{e.g.}, \cite{sutton2018reinforcement,deb2001multi,boehnlein2021artificial}). 
According to the DOE Town Halls on AI for Science \cite{stevens2020ai} these new approaches can ``\textit{revolutionize the way experimental nuclear physics is currently done}''. Recently it has been demonstrated by \cite{cisbani2020ai} that the Electron Ion Collider detector design can incredibly benefit from these techniques. 

\begin{figure}[!]
    \centering
    \includegraphics[trim=0 2cm 0 0, scale = 0.45]{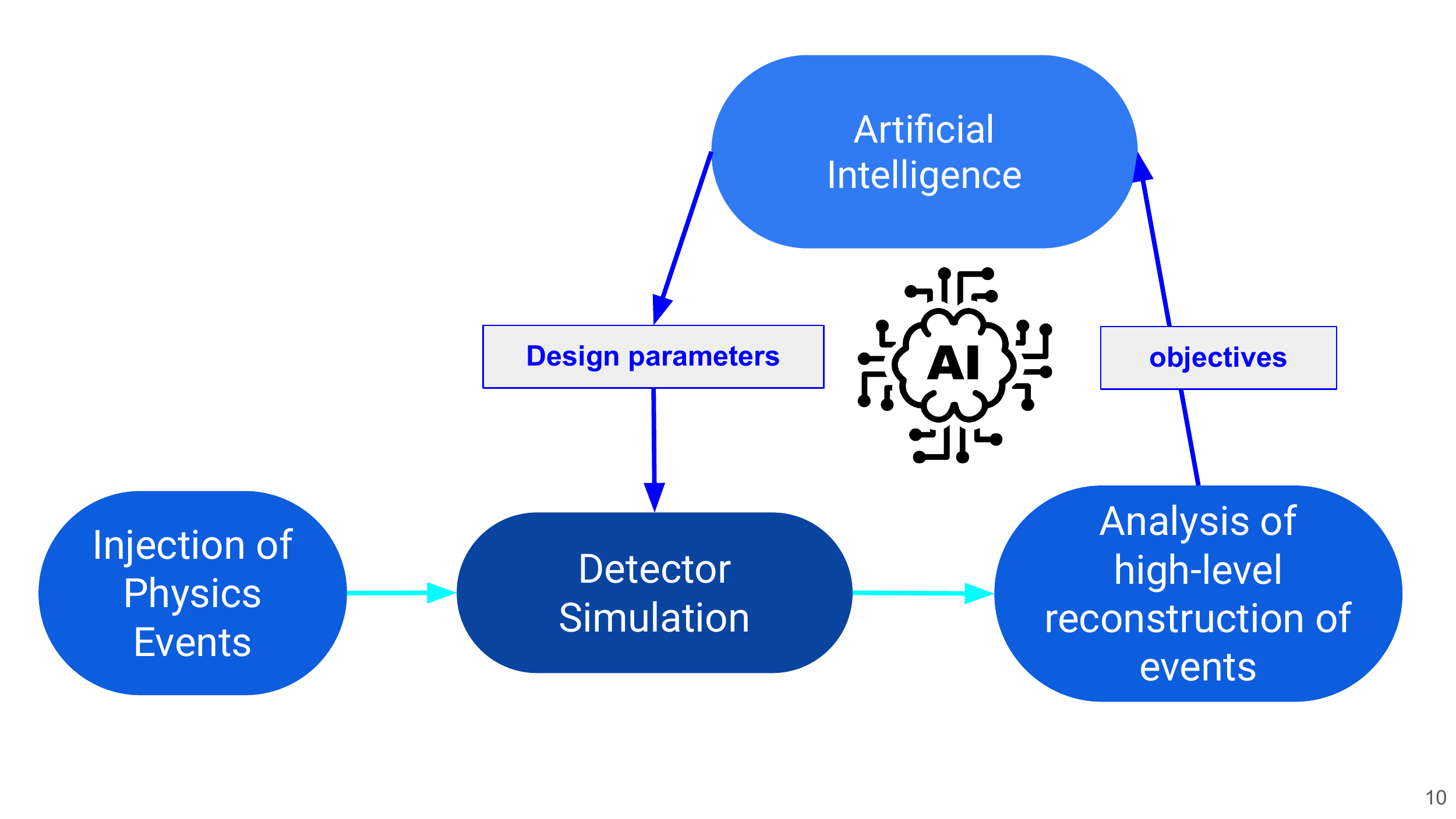}
    \includegraphics[trim=0 2cm 0 0, scale = 0.45]{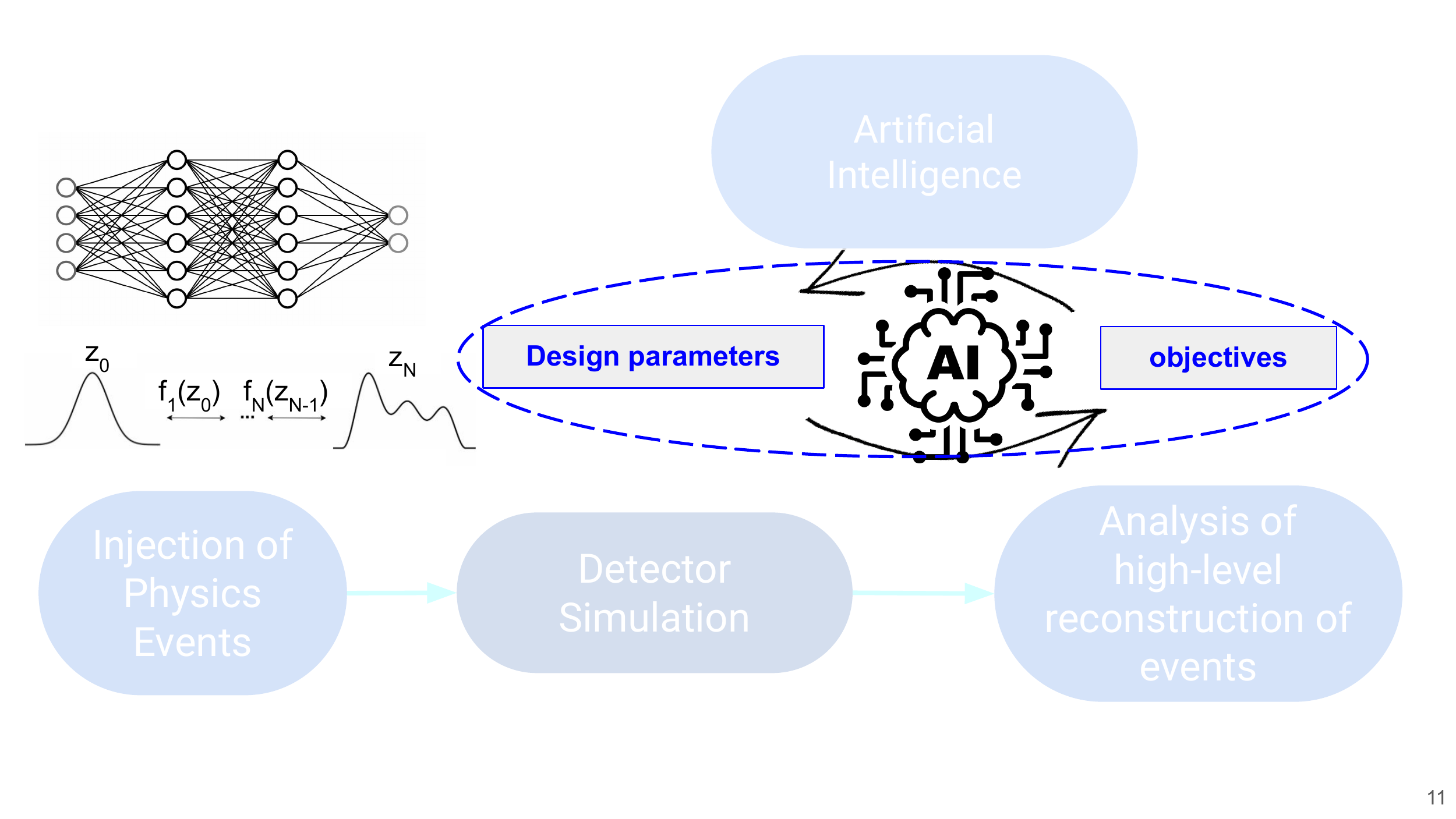}
    \caption{
    \textbf{Workflow of Detector Design:}
    (top) 
    Typical workflow of detector design assisted by AI: 
    physics events are injected in a detector characterized by some given design parameters. Reconstructed events are analyzed and figures of merit are quantified and passed to some AI-based strategy, which in turn suggests the next design point to observe in this sequential approach; notice that AI can also intervene in the simulation and reconstruction steps. 
    (bottom) With large dataset one can use deep learning to learn the `mapping' between the design and the objective space.  
    }
    \label{fig:design_AI}
\end{figure}

The typical workflow of detector design, represented in Figure \ref{fig:design_AI} (top), consists in the following steps: (i) physics events are generated and injected in a detector simulation (performed with Geant4) which depends on a vector of design parameters (\textit{design point}); (ii) events are reconstructed and high-level features are built and used as figures of merit which are passed to some AI-based strategy; (iii) AI in turn suggests the next design point to observe in this sequential approach. 
The reader should notice that in this workflow AI not only is used to assist the design process, but it can be actually incorporated also in the simulation and reconstruction steps, \textit{e.g.}, fast simulations based on generative adversarial networks (GANs) have been recently used by ATLAS to simulate hadrons in the calorimeter at intermediate energies \cite{atlas2021atlfast3}, and fast reconstruction methods based on variational autoencoders (VAEs) have been developed to identify particles with imaging Cherenkov detectors in \cite{fanelli2020deeprich}. 
In this regards, collaborations like MODE (an acronym for \textit{Machine-learning Optimized Design of Experiments} \cite{baydin2021toward}) aim at developing tools based on deep neural networks and modern automatic differentiation techniques to implement a full modeling of an experimental design, in order to achieve an end-to-end optimization of the design of instruments via a fully differentiable pipeline.
It should be pointed out though that while it is desirable to speed up the detector design pipeline in this way, CPU intensive simulations based on Geant4 make it difficult to produce large enough training datasets to take advantage of these techniques. Also, non-differentiable effects may arise from constraints that the detector design has to take into account during the optimization process. 

The remainder of this manuscript is organized as follows: in Sec.~\ref{sec:ecce_design}, a description of the approaches utilized to optimize the EIC detector design is provided along with results; in Sec.~\ref{sec:conclusions} perspectives and conclusions are given on how the design of detectors at the Electron Ion Collider with Artificial Intelligence can continue during the R\&D and design phases.



\section{AI Applications for the EIC Detector Design and R\&D}
\label{sec:ecce_design}

The first implementation of AI for optimizing the design of detectors in large-scale nuclear physics experiments has been realized for the dual-radiator Ring Imaging CHerenkov (dual-RICH) detector at the Electron Ion Collider \cite{cisbani2020ai}. 
This paper utilizes Bayesian Optimization (BO) \cite{jones1998efficient, snoek2012practical} --- which offers a derivative-free principled approach to global optimization of noisy and computationally expensive black-box functions --- and a single-objective  corresponding to the distinguishing power between pions ($\pi$) and kaons ($K$) --- which is the most challenging particle identification (PID) task for the dual-RICH considering that this detector has to provide a good $\pi/K$ separation over a wide range in momentum to meet the requirements of the EIC Yellow Report \cite{khalek2021science}. Gaussian processes (GP) have been used for regression, and a surrogate model has been reconstructed. Other ML-based regression approaches have been also utilized for comparison.
In \cite{cisbani2020ai}, BO with GP converged to the optimal solution with a minimal number of evaluations, though it should be pointed that differently from other approaches, the exact inference in GP regression is $\mathcal{O}$(n$^{3}$) where $n$ is the number of collected observations \cite{shahriari2015taking}, \textit{i.e.}, runtime grows as a function of the number of evaluations. 

\begin{figure}[!]
    \centering
    \includegraphics[scale = 0.5]{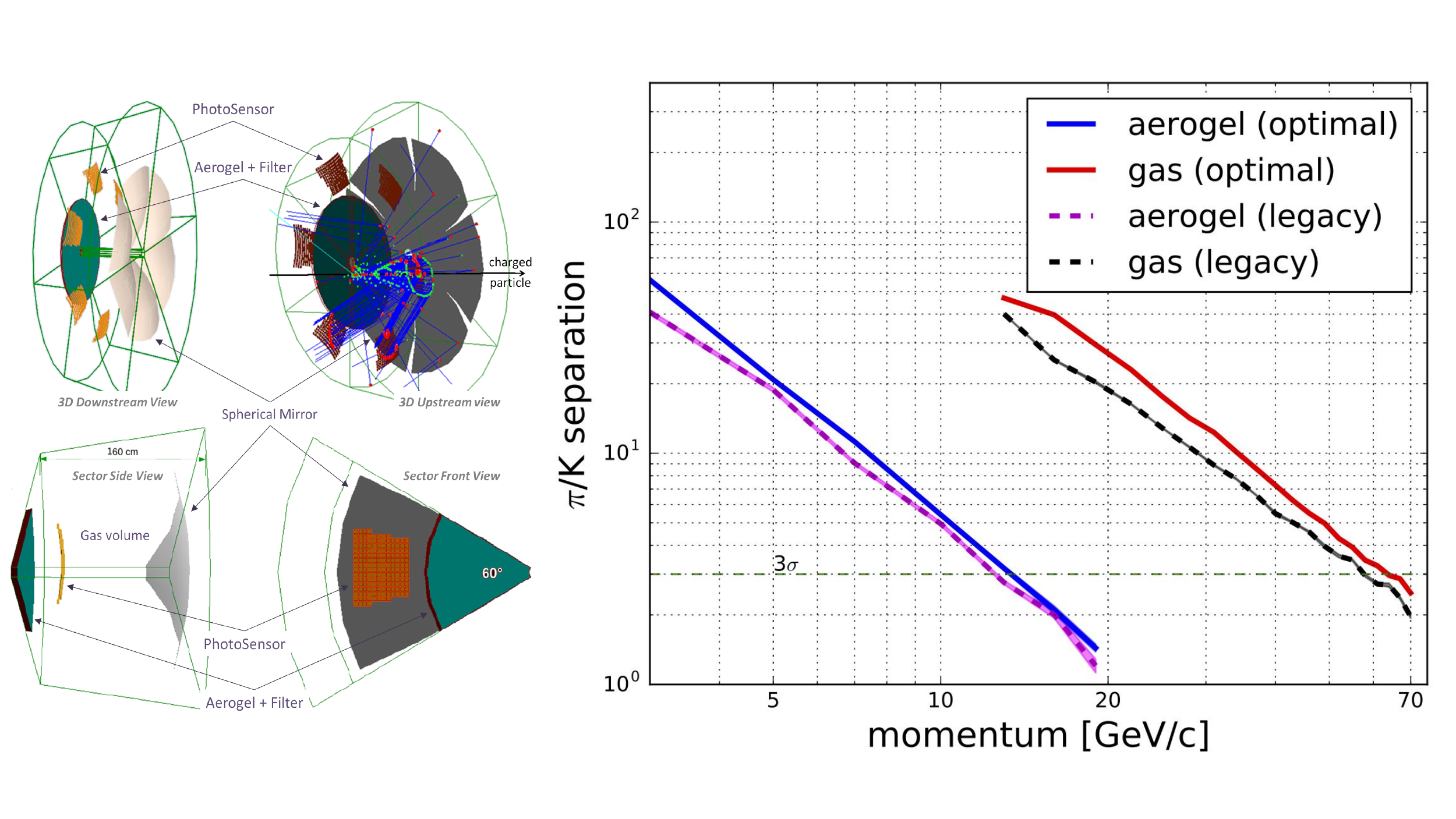}
    \includegraphics[trim=0 0.5cm 0 0, scale = 0.5]{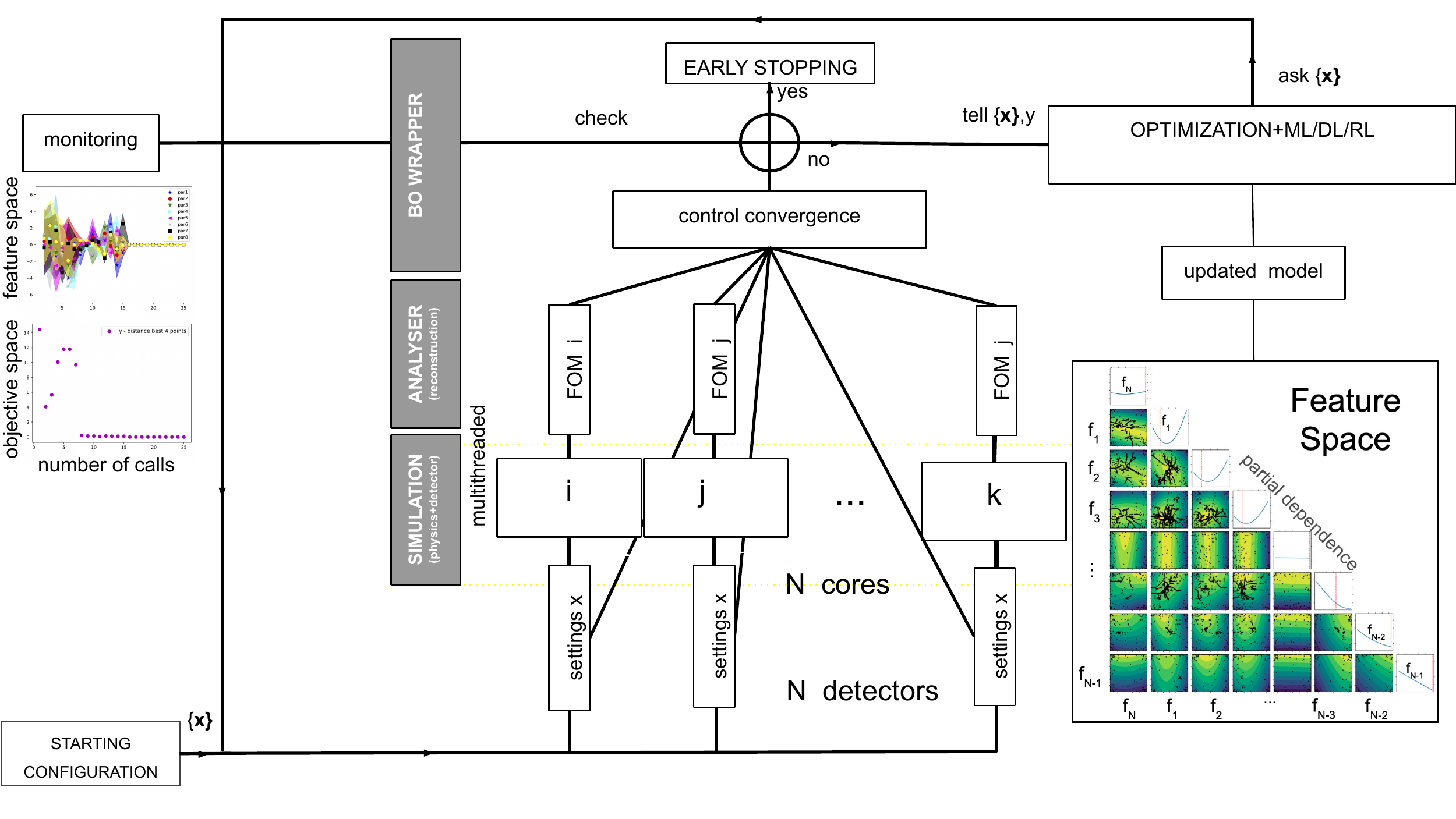}
    \caption{
    \textbf{AI-optimized dual-RICH design:}
    (top left): The dual-RICH design with an upstream and downstream view; a full description of the detector can be found in \cite{cisbani2020ai};
    (top right): $\pi / K$ separation as number of $\sigma$ as a function of the charged particle momentum. A statistically significant improvement is observed in the separation power with the approach discussed in this paper compared to the legacy baseline design. The curves are drawn with 68\% C.L. bands which are barely visible in the log plot, but this lets us better appreciate the significant difference between optimized and baseline curves. Figures are taken from \cite{cisbani2020ai};
    (bottom): parallelized and fully automated framework implemented in \cite{cisbani2020ai}. Convergence is monitored and activates early stopping criteria. 
    }
    \label{fig:dRICH}
\end{figure}

Fig.~\ref{fig:dRICH} shows in the top row the dual-RICH design and the $\pi/K$ separation power obtained by utilizing BO. 
The bottom row of Fig.~\ref{fig:dRICH} shows the automated,  highly-parallelized, and self-consistent workflow that has been developed by \cite{cisbani2020ai}. In total eight main design parameters have been identified that are capable to improve the PID performance of the dual-RICH: these are the refractive index and thickness of the aerogel radiator, the focusing mirror radius, its longitudinal and radial positions, and the three-dimensional  shifts of the photon sensor tiles with respect to the mirror center on a spherical surface. 
These studies resulted in a statistically significant improvement in the PID performance compared to an existing baseline design as shown in Fig.~\ref{fig:dRICH} (top right). 

As already discussed many design optimization problems are actually characterized by more than one objective. 
For the design of tracking system of the ECCE\footnote{The EIC Comprehensive Chromodynamics Experiment proto-collaboration.} detector, for example, a framework for Multi-Objective Optimization (MOO) has been embedded in the ECCE simulation software and utilized during the EIC detector proposal studies \cite{EIC_detector_proposal} to deal with a design problem parametrized as a multidimensional design space (D $\sim \mathcal{O}$(10), where D is the number of design parameters) and driven by multiple objectives such as the momentum resolution $dp/p$ of the tracks, their polar and azimuthal angular resolutions $d\theta, d\phi$ (with their projections to PID locations, \textit{e.g.}, the dual-RICH), along with the Kalman filter efficiency $\epsilon_{KF}$, while satisfying several mechanical constraints during the optimization process.\footnote{During the detector proposal we  utilized a minimum of 3 objectives and explored up to 4. The most effective set of objectives turned out to be $dp/p, d\theta, \epsilon_{KF}$. The other figures of merit have been evaluated in post hoc validation studies.  
More details can be found in \cite{fanelli2021ai, fanelli_ecce_tracker_ai}.}

Notice that MOO with dynamic/evolutionary algorithms (see, \textit{e.g.}, \cite{durillo2011jmetal,fortin2012deap, blank2020pymoo}) are probably the most utilized approaches, followed by multi-objective bayesian optimization (MOBO) which in the last few years has been characterized by a growing number of developments and applications (see, \textit{e.g.}, \cite{laumanns2002bayesian,balandat2019botorch,galuzio2020mobopt,mathern2021multi}), along with agent-based approaches like \cite{yang2019generalized}.\footnote{In accelerator physics there is an extensive use of MOO techniques, from MOEA like in \cite{yang2009global} for the optimization of the accelerator lattice to more recent works with MOBO as in \cite{roussel2021multiobjective} for accelerator tuning.} 
During the detector proposal studies, MOO has been used as an exploratory tool to steer the tracker design and help making decisions on the technology choice. 
Considering the complexity of the starting problem (D $\sim \mathcal{O}$(10), M=3,4) and the computing resources available during the detector proposal (each pipeline has been deployed on JLab sci-comp farm \cite{jlab_scicomp} utilizing one node with 128 CPU cores), meta-heuristic multi-objective genetic algorithms (MOGA) such as NSGA-II and NSGA-III \cite{deb2001multi, blank2020pymoo} have been used instead of principled approaches like MOBO because of some convenient aspects like (i) their intrinsic parallel nature, (ii) easier implementation of the analysis pipeline ($\textit{e.g.}$, does not require to identify suitable acquisition functions like in MOBO), and (iii) well known expected runtime complexity as a function of the population size N (the total number of design points generated at each iteration) and the number of objectives M, namely $\sim \mathcal{O}(MN^{2})$.\footnote{Notice that an improved version of NSGA-II scales as $\sim  \mathcal{O}(N \log{N^{(M-1)}})$ \cite{jensen2003reducing}.
 The main features of NSGA-II are (i) the usage of an elitist principle, (ii) an explicit diversity preserving mechanism, (iii) ability of determining non-dominated solutions; definitions of elitist principle, diversity and non-dominated solutions can be found in \cite{deb2001multi}.  
}
In principle NSGA is also suitable to be scaled on supercomputers as in \cite{liu2020parallelization} if an accurate reconstruction of the Pareto front \cite{debreu1954valuation} and therefore larger size populations are needed.

\begin{figure}[!]
    \centering
    \includegraphics[scale = 0.45]{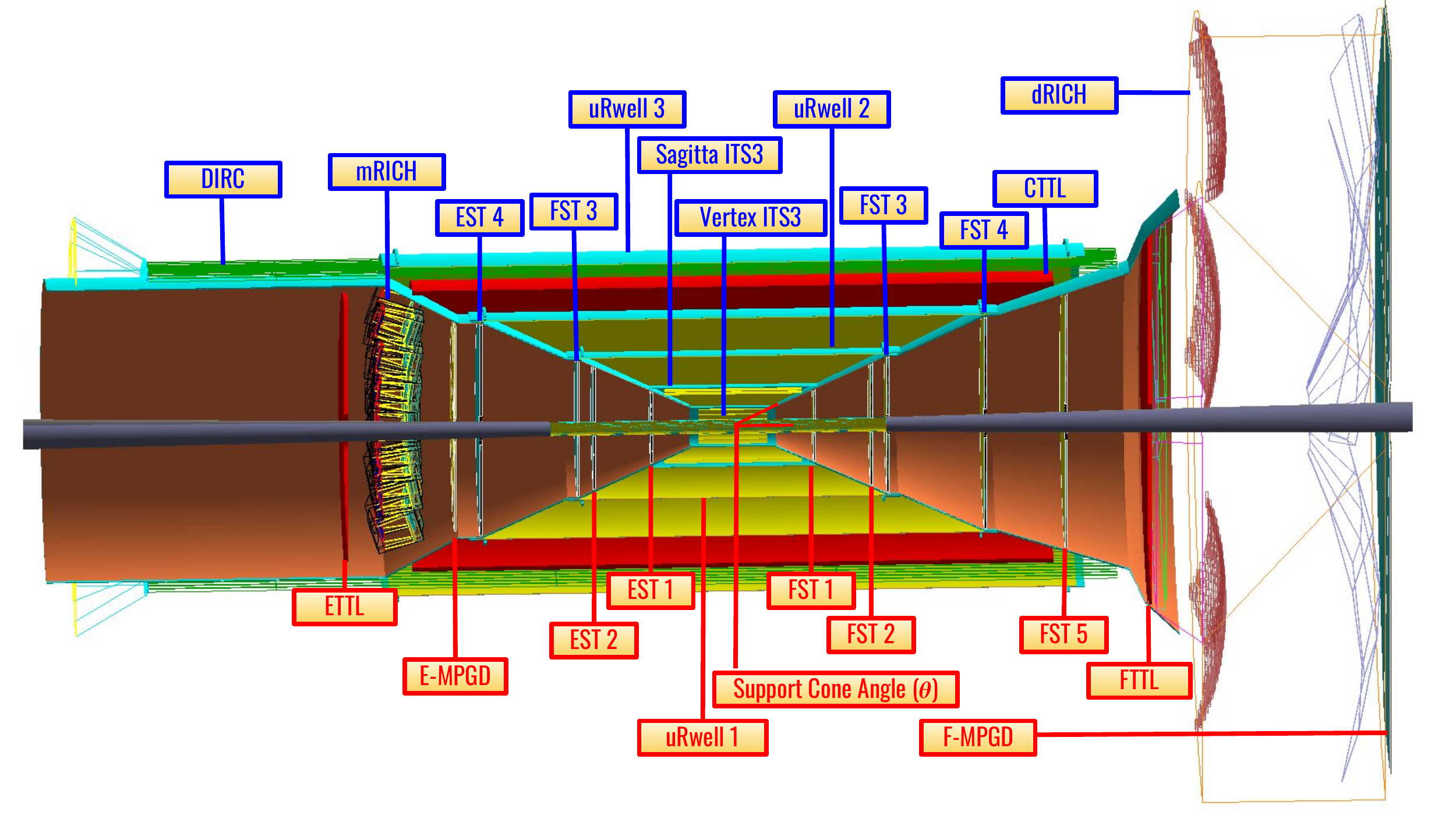}
    \includegraphics[scale = 0.45]{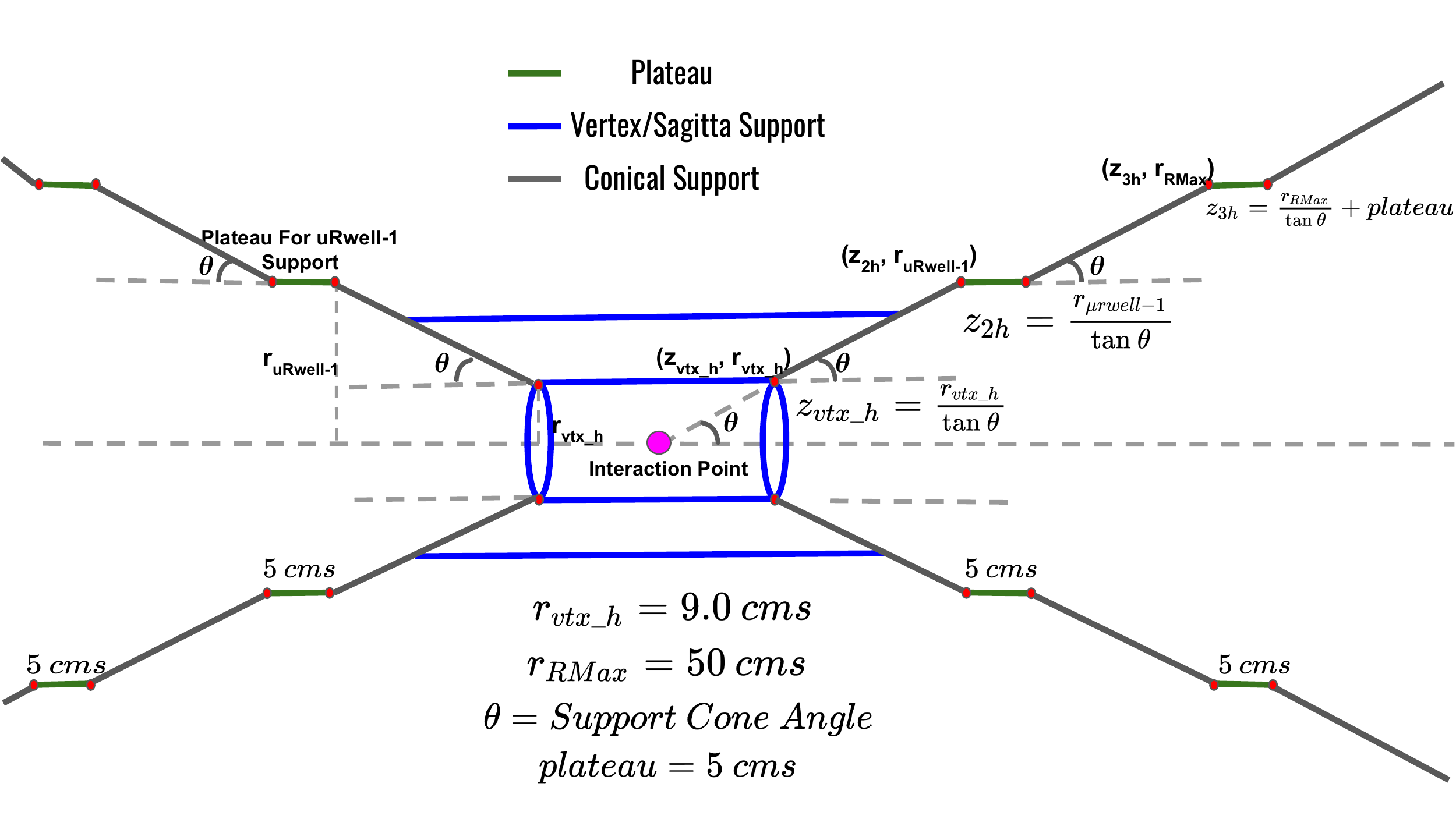}
    \caption{\textbf{ECCE tracking and PID system:}
    (top)  Starting from the interaction point, we have a conical support structure which defines the angle of the projective geometry.
    In the central (barrel) region we have 3 vertex barrel layers (ITS3 technology), followed by 2 sagitta layers (ITS3 technology), 2 $\mu$Rwell layers, ACLGAD based CTTL Time Of Flight (TOF) detector, DIRC (for PID) detector and finally another layer of $\mu$Rwell. In the electron-going direction there are 4 EST disks (silicon disks) followed by MPGD, mRICH (PID) detector and ACLGAD based ETTL TOF disk. In the hadron-going direction there are 5 FST disks (silicon disks) followed by a layer of ACLGAD based FTTL TOF disk, the dRICH (PID) detector and finally a layer of MPGD disk. The labels in red are the detector systems that are considered for optimisation. The labels in blue are the detector systems that are not free parameters and are not optimised. The tracking support cone comprises copper and plastic insulation along with carbon fiber and the cooling system (water).
    (bottom) Ongoing R\&D project with parametrization of the support cone for the inner tracker. The tracker support is characterised by 5 variables: $\theta$ (the angle of projection of the support cone structure), $r_{vtx}$ (radius of vertex support structure), $r_{\mu rwell-1}$ $\mu$Rwell-1 radius, plateau length, $r_{max}$ maximum allowed radius of inner tracker). More details can be found in \cite{fanelli_ecce_tracker_ai}.
    }
    \label{fig:SupportStructureParameterisation}
\end{figure}

During the detector proposal, many optimization pipelines have been run using M $\geq$ 3 objective functions, the total computing budget for an individual pipeline amounting to approximately 10k CPU-core hours. 
 Eventually the parametrization has been modified to include also the support structure in the design optimization problem. Overlaps in the design are checked before and during the optimization process and are excluded by the constraints and ranges of the parameters. The solutions obtained from the Pareto front are further checked to exclude overlaps.  
 Details on the ECCE tracker design and the parametrization of the ongoing project R\&D to optimize its support structure can be found in Fig.~\ref{fig:SupportStructureParameterisation}.
 
\begin{figure}[!]
    \centering
    \adjincludegraphics[trim=0 1.0cm 0 0, width=0.93\textwidth,Clip={.0\width} {.10\height} {0.\width} {.0\height}]{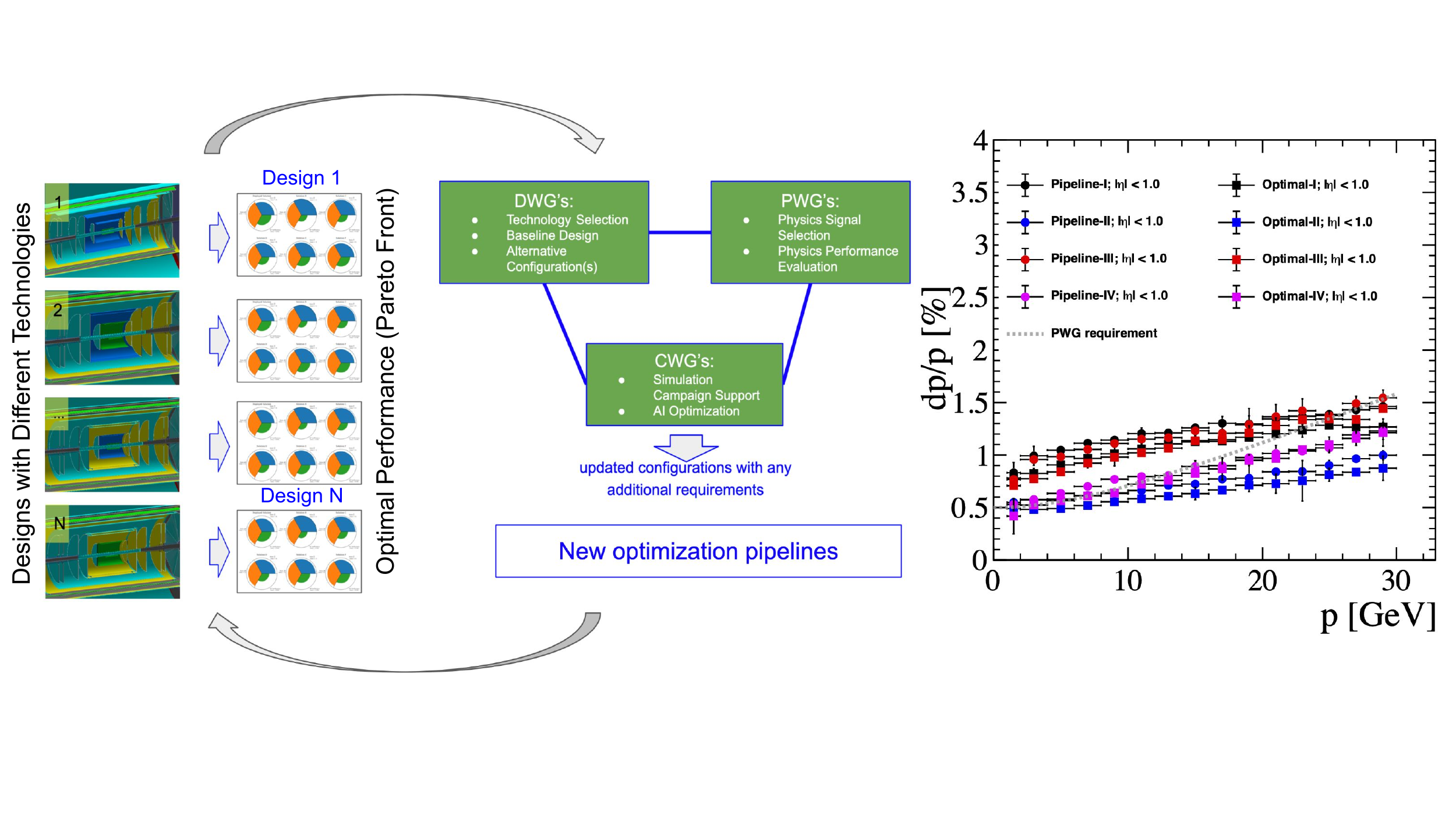}
    \caption{
    \textbf{Workflow describing the iterative process of the ECCE tracker design.} %
    This has been used during the detector technology choice to steer the design of the ECCE tracker. At a given instant in time, N alternative candidate configurations are studied. For each we create an optimization pipeline, which results in a Pareto front of solutions. The figure on the right shows for each pipeline (on the left) one of these tradeoff solutions and the corresponding optimized $dp/p$ curve in a central bin in pseudo-rapidity (performance must be evaluated in the entire phase-space and considering the other objectives too). 
    This new information helps the proto-collaboration steering the design: some configurations are rejected, while alternative ones are proposed to improve the design. New optimization pipelines are defined and the process is iterated. 
    The fundamental interplay between the ECCE teams working on Physics (PWGs), Detector (DWGs) and Computing (CWGs) is thus propelled by AI during the design process. More details can be found in \cite{fanelli_ecce_tracker_ai}.
}
    \label{fig:pipeline_workflow}
\end{figure}

AI has played a crucial role in helping to choose a combination of technologies for the ECCE inner tracker and has been used as input to multiple iterations of the ECCE tracker design, leading to the current tracker layout. As shown in Fig.~\ref{fig:pipeline_workflow}, this is a continued optimization process that evolved in time and required the interplay between the ECCE teams working on Physics, Detector and Computing. Results are validated by looking at figures of merit that do not enter explicitly as objective functions in the optimization process (\textit{e.g.}, physics-driven objectives as explained in what follows). The decision making is left post hoc and discussed among the ECCE teams.
 The tracker design has been optimized using particle gun samples of pions; after the optimization Pythia \cite{sjostrand2020pythia} samples have been used to evaluate, \textit{e.g.}, the improvement in the reconstruction of ${D}^{0}$ meson decays into $\pi^{+}K^{-}$ from semi-inclusive deep inelastic scattering events.\footnote{Similarly other physics analyses are eventually run utilizing the new proposed design and comparisons are made with results obtained with previous designs.}

\begin{figure}[b]
    \centering
    \vspace{-0.cm}
    \includegraphics[scale = 0.43]{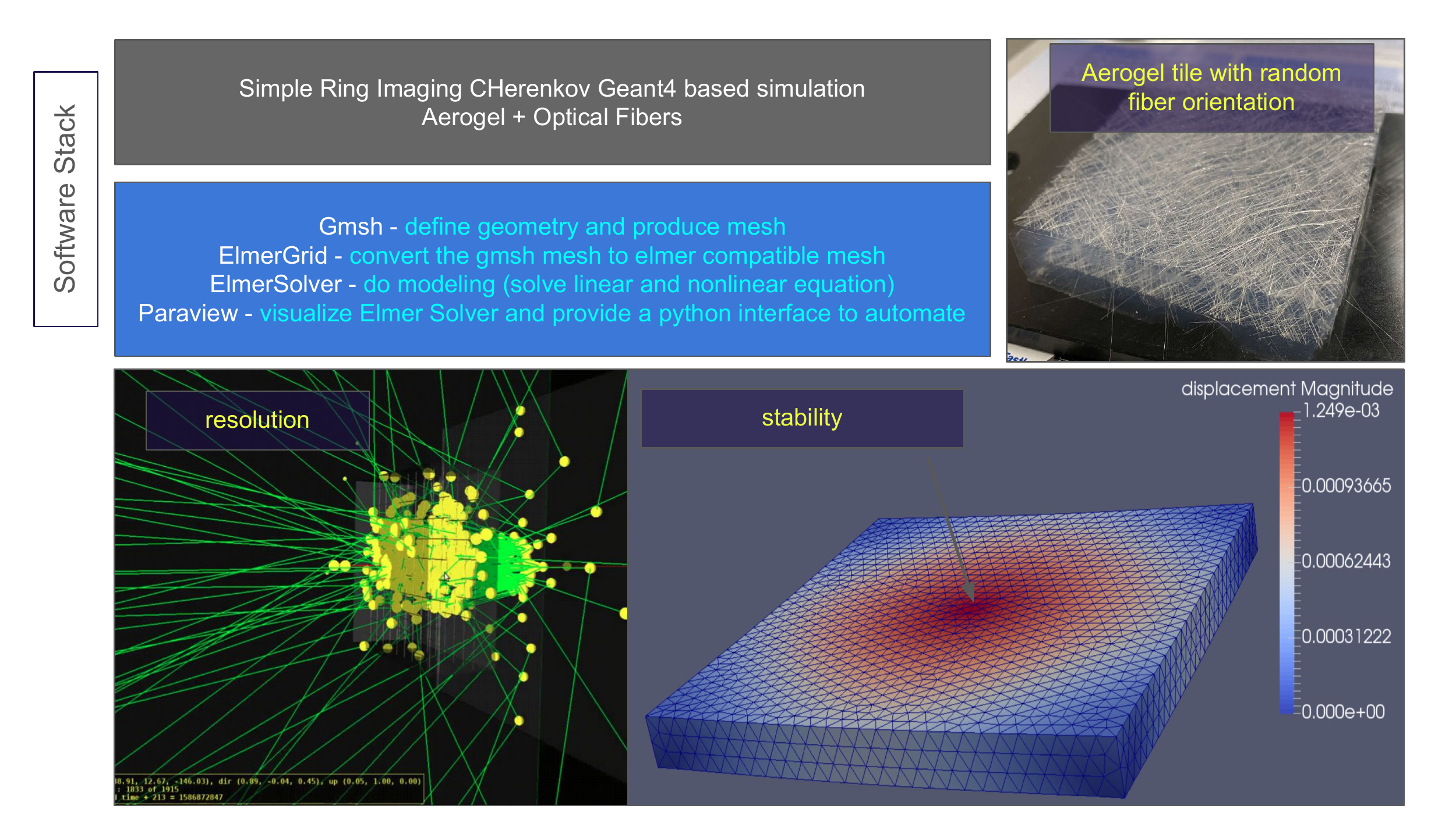}
    \caption{
    \textbf{Reinforced novel aerogel material with fibers: \textit{aerfi}}. Aerogel tiles can be reinforced with structured meshes of fibers. From top left in clockwise order: the software stack necessary to simulate the Cherenkov angle resolution and the mechanical stability of the fibers; an experimental realization of reinforced aerogel tile with a random orientation of the fibers; a visualization of the mechanical stability studies; a visualization of the Geant4 simulation of an aerogel tile with fibers.
    }
    \label{fig:aerfi}
\end{figure}

 AI provided insights on the design of the ECCE tracker, showing how much the tracking system with different technology choices could be optimized. An example of these improvements is represented in Fig. \ref{fig:pipeline_workflow} (right). 
 Ultimately this continued AI-assisted optimization process brought to a projective design. 
 A two-level parallelization has been implemented in the MOO framework, the first creating the parallel simulations of design points, the second parallelizing each design point. The evaluation itself can be distributed to several workers or a whole cluster with libraries like Dask \cite{rocklin2015dask}.
Computing time studies have been also carried out in \cite{fanelli_ecce_tracker_ai} to evaluate the simulation time of each single design point as a function of the number of tracks generated as well as to characterize the computing time taken by the AI-based algorithm in generating a new population of design points. 
In the current implementation the simulation time dominates the time taken by NSGA to update the population of design points and a better tradeoff can be found with more computing resources made available to support an improved parallelization strategy. 
It turned out that a population size N=100 and offspring O=30 was sufficient to reach a good approximation of the Pareto front for a 10-dimensional problem with 3 objectives. Again larger size populations can allow to approximate the Pareto front with better accuracy.
A larger population size may actually become necessary for fine-tuning the final detector concept once the optimization problem is extended to simultaneously optimize multiple systems of sub-detectors (\textit{e.g.}, tracker, PID detectors, etc).


MOGA techniques are also being currently investigated in ongoing R\&D efforts for developing new aerogel materials that can be utilized in experiments like EIC where Imaging Cherenkov detectors constitute the backbone of PID \cite{cfanelli_ai4eic_aiml_design}, and the dual-RICH will utilize two radiators, one of these being aerogel-based.

 Aerogels tiles with low refractive indices tend to be very fragile and can break during production and handling, or during the installation in detectors.
To improve the mechanical strength of aerogels a reinforcement strategy has been developed: the general concept consists in introducing fibers into the aerogel that increase the robustness and stability of the tile which can be thought of as a pseudo Young's modulus but that at the same time do not affect the optical properties of the aerogel, the latter quantified by the resolution of the Cherenkov angle. 
These two figures of merit are competing with each other (the larger the number of the fibers the poorer the resolution, and we want to both maximize the stability but at the same time improve the resolution) and therefore we are interested in identifying the Pareto front in this two-dimensional objective space with MOO techniques.  
The software stack, also represented in Fig.~\ref{fig:aerfi}, includes Geant4 simulations \cite{agostinelli2003geant4} to simulate a simple ring imaging Cherenkov with aerogel and fibers, and a combination of (i) Gmsh \cite{geuzaine2009gmsh}, to define the geometry and produce the fibers mesh, (ii) ElmerGrid \cite{raaback2015elmergrid}, to convert the gmsh mesh to elmer compatible mesh, (iii) ElmerSolver \cite{ruokolainen2016elmersolver}, to do the modeling and solve linear and nonlinear equations and (iv) Paraview \cite{ahrens2005paraview}, to visualize the results of ElmerSolver and provide a python interface for automatization.

Fig.~\ref{fig:aerfi} includes also the image of an experimental realization of a reinforced aerogel tile with randomized orientation of the fibers, and visualizations of the Gmsh+Elmer and Geant4 simulations, respectively.
Work is ongoing to validate these simulations by comparing the results with experimental data.

\section{Conclusions and Perspectives}
\label{sec:conclusions}

AI has been successfully utilized at EIC for the design optimization of the dual-RICH (tackled as a single objective optimization problem) \cite{cisbani2020ai} and more recently for the multi-objective optimization of the ECCE tracker system \cite{fanelli2021ai} during the EIC detector proposal studies. 
MOO is also currently being used by a R\&D project to design novel aerogel material that may be utilized for PID. 
AI will continue to be utilized in the next few years to further optimize the design of the EIC detector, 
 after the final detector concept is chosen. 
We have seen how these AI-based strategies are particularly useful for the design optimization of large-scale experiments like EIC in that compute intensive simulations, a high-dimensional design parameter space and multiple objectives are involved in this process.

Utilization of ML/DL for fast simulations is desirable but typically complicated by the need of large datasets which have to be produced as a function of different design points for the exploration of the optimal design solutions. Therefore Geant4 simulations at least in a first phase seem to be necessary. 
Interestingly the eAST (eA simulation toolkit, \cite{project_east}) project is aiming at implementing a \textit{common and integrated approach for fast and full detector simulations in Geant4 with a plug and play modular approach} which \textit{will ensure that any work on AI/ML running on computers with heterogeneous hardware configurations can be directly applied}.

Once the selection of the EIC collaboration is made, these AI-based strategies can help to further optimize the proposed detector concept based on the experience of ECCE during the detector proposal.
The design optimization can be extended to include a larger system of tunable sub-detectors, taking into account the constraints from the global detector design.  
Physics analyses are at the moment done after the optimization for a given detector design solution candidate, but we aim to encode  during the optimization process physics-driven objectives in addition to objectives representing the detector performance.
A thorough comparison of results obtained with different AI-based strategies (\textit{e.g.}, MOGA, MOBO) will be also studied.
As already discussed, another interesting avenue to explore in the future is the deployment of MOO approaches on exascale supercomputers for larger parametrizations of the global design. 

In conclusion, the design of large scale experiments in high energy nuclear physics like the future EIC involve unprecedented computational challenges and the optimization of such complex systems can benefit from state of the art AI-based strategies.     
The Electron Ion Collider is one of the first experiments to systematically leverage Artificial Intelligence starting from the design and R\&D phases.

\acknowledgments

The work of CF is supported by the U.S. Department of Energy, Office of Science, Office of Nuclear Physics under grant No. DE-SC0019999.




\bibliographystyle{unsrturl}
\bibliography{refs}

\end{document}